\documentclass[twocolumn,showpacs,aps,prl]{revtex4}
\usepackage{graphicx}
\usepackage{dcolumn}
\usepackage{amsmath}

\begin{document}

\title{
Identified Particle Elliptic Flow in Au+Au Collisions at 
$\sqrt{s_{_{NN}}}=130$~GeV}

\author{
C.~Adler$^{11}$, Z.~Ahammed$^{23}$, C.~Allgower$^{12}$, J.~Amonett$^{14}$,
B.D.~Anderson$^{14}$, M.~Anderson$^5$, G.S.~Averichev$^{9}$, 
J.~Balewski$^{12}$, O.~Barannikova$^{9,23}$, L.S.~Barnby$^{14}$
% ...
\begin{center}(STAR Collaboration)\end{center}
}

\affiliation{
% ...
$^5$University of California, Davis, California 95616 \\
% ...
$^{9}$Laboratory for High Energy (JINR), Dubna, Russia \\
% ...
$^{11}$University of Frankfurt, Frankfurt, Germany \\
$^{12}$Indiana University, Bloomington, Indiana 47408 \\
% ...
$^{14}$Kent State University, Kent, Ohio 44242 \\
% ...
$^{23}$Purdue University, West Lafayette, Indiana 47907 \\
% ...
}

\begin{abstract}
We report first results on elliptic flow of identified particles at
mid-rapidity in Au+Au collisions at $\sqrt{s_{_{NN}}}=130$~GeV using
the STAR TPC at RHIC\@. The elliptic flow as a function of transverse
momentum and
centrality differs significantly for particles of different masses.
This dependence can be accounted for in hydrodynamic models,
indicating that the system created shows a behavior consistent with
collective hydrodynamical flow. The fit to the data with a simple
model gives information on the temperature and flow velocities at
freeze-out.

\end{abstract}


\maketitle

The goal of the ultra-relativistic nuclear collision program is the
creation of a system of deconfined quarks and gluons~\cite{qm99}. The
azimuthal anisotropy of the transverse momentum distribution for
non-central collisions is thought to be sensitive to the
early evolution of the system~\cite{sorge}. 
The second Fourier coefficient of this
anisotropy, $v_{2}$, is called elliptic flow. It is an important
observable since it is sensitive to the rescattering of the
constituents in the created hot and dense matter.  This rescattering
converts the initial spatial anisotropy, due to the almond shape of
the overlap region of non-central collisions, into momentum
anisotropy.  The spatial anisotropy is largest early in the
evolution of the collision, but as the system expands and
becomes more spherical, this driving force quenches itself. Therefore,
the magnitude of the observed elliptic flow should reflect the extent 
of the rescattering at relatively early time~\cite{sorge}.

Elliptic flow in ultra-relativistic nuclear collisions was first
discussed in Ref.~\cite{olli92} and has been studied intensively in
recent years at the Alternating Gradient Synchrotron 
(AGS)~\cite{e877flow2,e895},
Super Proton Synchrotron (SPS)~\cite{na49prl,na49flow,wa98} and at the
Relativistic Heavy Ion Collider (RHIC)~\cite{starflow} energies.
The studies at top AGS energy and SPS have found that elliptic flow at
these energies is in the plane defined by the beam direction and 
the impact parameter, $v_2>0$, as expected from most models. 
The pion elliptic flow for relatively peripheral collisions increases 
with beam energy~\cite{olli98} from about 0.02 at the top AGS
energy~\cite{e877flow2}, 0.035 at the SPS~\cite{na49flow} to about 0.06
at RHIC energies~\cite{starflow}. The increased magnitude of the
integrated elliptic flow at RHIC reaches the values predicted by
hydrodynamical models, which are based on the assumption of complete 
local thermalization.

The first elliptic flow results from RHIC were for charged particles.
The differential charged particle flow, $v_2$($p_t$), shows an almost
linear rise with transverse momentum, $p_t$, up to 1.5 GeV/$c$. 
At $p_t >$ 1.5 GeV/$c$, the $v_2$($p_t$) values start to saturate, 
which might indicate
the onset of hard processes~\cite{starflow,wang,gyulassy,molnar}.
The behavior of $v_2$($p_t$) up to 1.5 GeV/$c$ is consistent with a
hydrodynamic picture.
Hydrodynamics assumes complete local thermalization at the formation
of the system, followed by an evolution governed by an Equation Of
State (EOS). However, the $p_t$-integrated elliptic flow, $v_2$, as a
function of centrality, and the differential $v_2$($p_t$), show little
sensitivity to the EOS used~\cite{kolb1}.
Studies of the mass dependences of elliptic flow
for particles with $p_t <$ 1.5 GeV/$c$ provide important additional
tests of the hydrodynamical model~\cite{pasi2}. 
Similar to the identified single
particle spectra, where the transverse flow velocity has been
extracted from the mass dependence of the slope parameter~\cite{nxu},
the $v_2$($p_t$) for different mass particles allows the extraction of
the elliptic component of the flow velocity~\cite{eos,voloshin_e877}. 
Moreover, the details of the dependence of elliptic flow on particle 
mass and transverse momentum are sensitive to 
the temperature, transverse flow
velocity, its azimuthal variation, and source deformation
at freeze-out. In this Letter we report the first results for
identified particle $v_2$($m,p_t$) in Au+Au collisions at RHIC
at $\sqrt{s_{_{NN}}}=130$~GeV. 

The Solenoidal Tracker At RHIC (STAR)~\cite{STAR}, 
is ideally suited for measuring elliptic flow due to its
azimuthal symmetry and large coverage. The detector
consists of several sub-systems in a large solenoidal magnet. The Time
Projection Chamber (TPC) covers the pseudorapidity range $|\eta| <
1.8$ for collisions in the center of the TPC\@. The magnet was operated
at a 0.25 Tesla field, allowing tracking of particles with
$p_t>75$~MeV/c. Two Zero Degree Calorimeters~\cite{ZDC} located at
$|\sin\theta| <$ 0.002, which mainly detect fragmentation neutrons, are
used in coincidence for the trigger. The TPC is surrounded by a
scintillator barrel which measures the charged particle multiplicity
within $|\eta| < 1$, for triggering purposes.

For this analysis, 120,000 minimum-bias events were selected with a 
primary vertex
position within 75~cm longitudinally of the TPC center and within 1~cm
radially of the beam line. For determination of the 
event plane~\cite{starflow},
charged particle tracks were selected with $0.1 < p_t \le 2.0$ GeV/c.
All tracks used in this analysis passed within 2~cm of the primary
vertex and had at least 15 measured space points. 
Also, the ratio of the number of measured space
points to the expected maximum number of space points for that
particular track was required to be greater than 0.52, 
suppressing split tracks from being counted more than once. 
The tracks used for the determination of the reaction plane 
were within $|\eta| < 1.0$, and the tracks used to calculate the
elliptic flow were within $|\eta| < 1.3$. 
These cuts are similar to the ones used in
Ref.~\cite{starflow}, and the analysis results presented here are
not sensitive to those cuts.

\begin{figure}[t]
  \centering\mbox{
    \includegraphics[width=.49\textwidth]{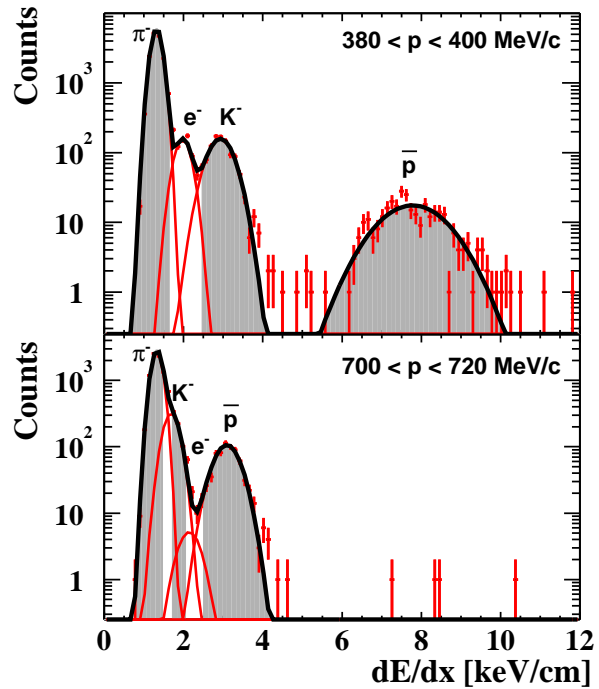}}
  \caption{Charged particle identification for central events
    at $|\eta| \le 0.1$ for the
    momentum range of 380 -- 400 MeV/c and 700 -- 720 MeV/c.
    The shaded areas show the selected ranges for the different
    particles.}
  \label{dedx}
\end{figure}

The pions, protons, and antiprotons were selected according to
their specific energy loss ($dE/dx$) in the TPC in the transverse 
momentum range of
0.175 -- 0.75 GeV/$c$, 0.5 -- 0.9 GeV/$c$ and 0.3 -- 0.9 GeV/$c$,
respectively. The lower $p_t$ values were chosen such that the energy
loss in the detector was negligible. 
At low momentum, the proton background due to secondary interactions
with the detector material is significant. Therefore, only 
protons above a transverse momentum of 0.5 GeV/$c$ were used in this
analysis. At a momentum ($p$)
of 0.5 GeV/$c$, the $dE/dx$ resolution was found to be of the order of
11\% for a typical long track in the STAR TPC\@. The kaons, because of
their overlap in $dE/dx$ with the electrons and positrons, were
selected in the momentum ranges of 0.30 -- 0.40 GeV/$c$ and 0.60 --
0.70 GeV/$c$. The raw yields of the pions, kaons, protons, and
antiprotons were obtained from fitting the $dE/dx$ distributions for
each $\eta$, $p$ and centrality bin with multiple Gaussians, 
and requiring
greater than 90\% purity (particle's yield divided by the sum of
yields for all particles at that same $dE/dx$). 
Figure~\ref{dedx} shows the $dE/dx$
distributions for negative tracks in the TPC for central events and
$|\eta| \le 0.1$. The upper panel shows the $dE/dx$ distribution for
the momentum range of 0.38 -- 0.40 GeV/$c$ and the multiple Gaussian
fit. In this momentum range, the $\pi^{-}$, $e^{-}$, K$^{-}$ and
$\bar{p}$ are clearly separated.  The lower panel shows the $dE/dx$
distribution for the momentum range of 0.70 -- 0.72 GeV/$c$. For the
higher momentum bin, the $\pi^{-}$, $e^{-}$ and K$^{-}$ overlap;
however, the $\bar{p}$ is still clearly separated. Even though it is
still possible to extract kaons with 90\% purity in this momentum bin,
we only used the kaons up to 0.70 GeV/$c$.

The flow analysis method~\cite{meth} involves the calculation of the
event plane angle, which is an experimental estimator of the real
reaction-plane angle. For the analysis presented in this Letter,
each particle was correlated with the event plane from all the other
particles (for the other methods, see~\cite{QMsnellings}).
\begin{figure}[t]
  \centering\mbox{
    \includegraphics[width=.49\textwidth]{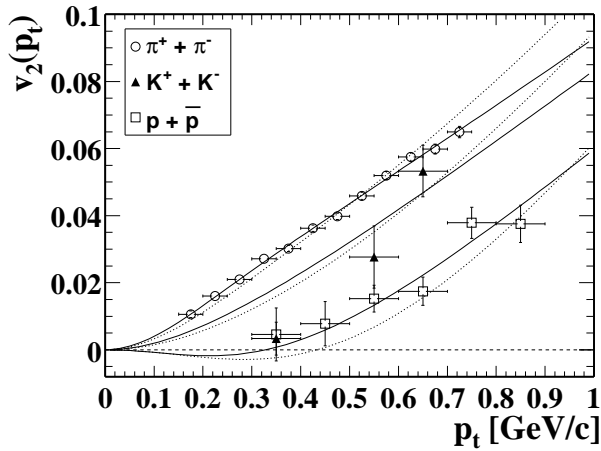}}
  \caption{Differential elliptic flow for pions, kaons and protons 
    + antiprotons for minimum-bias events.
    The solid lines show the fit with the modified Blast Wave model,
    and the dotted lines are the fit with the unmodified model.}
  \label{pidflow1}
\end{figure}
The differential elliptic flow, $v_2$, depends on mass, rapidity ($y$) and
$p_t$. In Fig.~\ref{pidflow1}, $v_2$($p_t$) is shown for pions, kaons,
and protons + antiprotons for minimum-bias
collisions~\cite{footminbias}, integrated over rapidity and centrality
by taking 
the multiplicity-weighted average. 
The uncertainties shown are statistical only. 
Using the same procedure to estimate the systematic uncertainties as in
Ref.~\cite{starflow} for $v_2$ integrated over $p_t$, 
the systematic uncertainty for minimum-bias data is 13\%.
We have verified that the positive and
negative identified particles used in this analysis have the same
$v_2$($p_t$) within statistical uncertainties. For $v_2$($p_t$), the pions
were integrated over $|y| \le 1.0$, the kaons over $|y| \le 0.8$, and
the protons + antiprotons over $|y| \le 0.5$ (the rapidity ranges chosen
correspond to approximately consistent $|\eta|$ coverage for all of the 
selected particles). 
Mathematically the $v_2$ value at $p_t = 0$, as well as its first
derivative, must be zero.
As a function of $p_t$, the pions exhibit an
almost linear dependence of $v_2$, whereas 
the protons + antiprotons exhibit a more quadratic behavior, clearly
different from the pions. As expected in a hydrodynamic picture, the
kaons lie between the pions and the protons + antiprotons.
The 90\% purity of the protons + antiprotons in the $0.8 \le p_t \le
0.9$ range leads to a maximum +5\% systematic error on the $v_2$ value
in this bin.
The observed behavior may be the result of the interplay between the mean 
expansion
velocity, the elliptic component of the expansion velocity, and the
thermal velocity of the particles.
A similar effect was
discussed for the case of directed flow in~\cite{voloshinv1}.

\begin{figure}[t]
  \centering\mbox{
    \includegraphics[width=.49\textwidth]{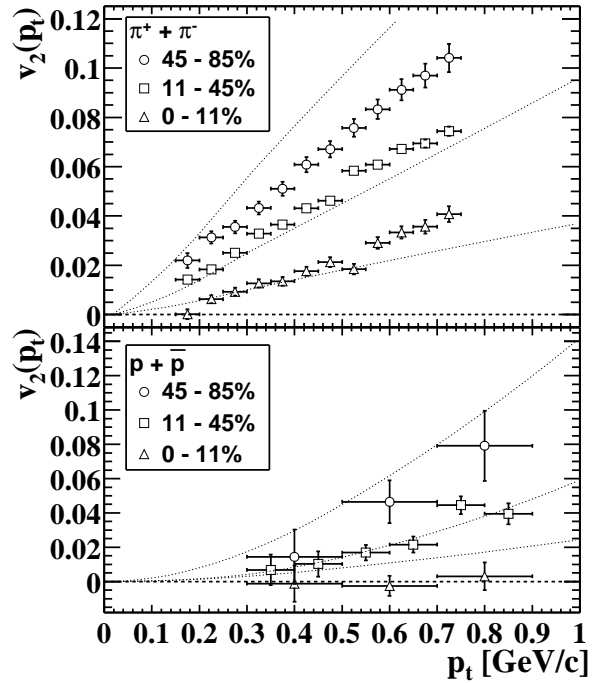}}
  \caption{Upper panel: Differential elliptic flow for pions in three 
    different centrality bins. 
    Lower panel: The same for protons + antiprotons.
    The dotted lines show the
    predictions from a full hydrodynamic model
    calculation~\cite{pasi2}.
    The uncertainties shown are statistical only.}
  \label{pidflow2}
\end{figure}

The differential elliptic flow, $v_2$($p_t$), is plotted
for pions for three different centrality selections in the upper panel
of Fig.~\ref{pidflow2}, and for protons in the lower panel. The 
open triangles represent the most central 11\% of the
measured cross section. The open squares
correspond to 11 -- 45\% of the measured cross section, and the open
circles correspond to 45 -- 85\%. The
uncertainties on the points are statistical only. The systematic 
uncertainty is smallest for the centrality region with the best
reaction plane resolution and is
estimated to be 20\% for the most central bin, 8\% for the mid-central
bin, and 22\% for the most peripheral bin. At a given $p_t$, the more
peripheral collisions have the largest value of $v_2$($p_t$), and 
$v_2$($p_t$)
decreases for more central collisions. 
For all three of the centrality ranges, in the measured $p_t$ range, 
the $p_t$ dependence of $v_2$ for pions is approximately linear.

We have fitted the data with a simple hydrodynamic-motivated model. 
This model is a generalization of the Blast Wave model 
from~\cite{Siemens,pasi2} assuming the flow field is perpendicular to the
freeze-out hyper-surface. 
\begin{equation} 
  v_2(p_t)= 
  \frac{  \int_0^{2\pi} d\phi_b 
    \cos(2 \phi_b) 
    I_2(\alpha_t) K_1 (\beta_t) 
    (1+2 s_2 \cos(2\phi_b))} 
  {  \int_0^{2\pi} d\phi_b
    I_0(\alpha_t) K_1 (\beta_t) 
    (1+2 s_2 \cos(2\phi_b))}, 
  \label{modblastwave}
\end{equation} 
where $I_0$, $I_2$, and $K_1$ are modified Bessel functions, and where
$\alpha_t(\phi_b)=(p_t/T_f)\sinh(\rho(\phi_b))$, and 
$\beta_t(\phi_b)=(m_t/T_f)\cosh(\rho(\phi_b))$. 
The basic assumptions of this model are 
boost-invariant longitudinal expansion and freeze-out at constant 
temperature $T_f$ on a thin shell, which expands with a 
transverse rapidity exhibiting a second harmonic azimuthal modulation 
given by $\rho(\phi_b)=\rho_0 +\rho_a \cos(2 \phi_b)$. 
In this equation, $\phi_b$ is the azimuthal angle (measured with 
respect to the reaction plane) of the boost of the source element 
on the freeze-out hyper-surface~\cite{pasi2}, and $\rho_0$ and
$\rho_a$ are the 
mean transverse expansion rapidity ($v_0 = \tanh(\rho_0)$) and the 
amplitude of its azimuthal variation, respectively. 
With $s_2 = 0$, our equation reduces to Eq.2 from~\cite{pasi2}. 
In Fig.~\ref{pidflow1}, the fit to the minimum-bias data with $s_2 = 0$ 
is shown as the dotted lines. 
The poor fit shows that the data can not be described under 
the assumption of a spatially isotropic freeze-out 
hyper-surface in the transverse plane. 
This led us to generalize Eq.2 from~\cite{pasi2} to the 
case of a spatially anisotropic freeze-out hyper-surface, 
introducing one extra 
parameter, $s_2$, describing the variation in the azimuthal density of 
the source elements, $\propto 2 s_2 \cos(2 \phi_b)$. 
This additional parameter leads to a good description of the data, 
shown as the solid lines in Fig.~\ref{pidflow1}. 
A positive value of the $s_2$ parameter would mean that there are more
source elements moving in the direction of the reaction plane.
The $s_2$ parameter does not distinguish between an in-plane or
out-of-plane extended source, because the direction of the flow field
is not an observable. However, azimuthally sensitive HBT measurements
will be able to address this. 
In this model (Eq.~\ref{modblastwave}), elliptic flow as a function of 
particle transverse momentum, $v_2$($p_t$), starts from zero and rises 
approximately quadratically with $p_t$ until the particle becomes 
relativistic, and then  $v_2$($p_t$) continues to rise almost linearly. 
For the heavier particles, the linear rise is delayed. 
These mass-dependent effects are larger for lower temperatures 
($T_f$) and larger transverse rapidities ($\rho_0$). The linear rise 
increases with both the amplitude of the azimuthal variation of the 
transverse expansion rapidity ($\rho_a$) and the 
elliptic deformation ($s_2$), but only $\rho_a$ produces a mass 
dependence. This arises from the dependence of the $s_2$ parameter
effect on momentum not energy.
This also results in a change in slope and a characteristic curvature in 
$v_2$($p_t$), since the flow associated with $s_2$ saturates quite early, 
when $p_t/T \gg 1$~\cite{pasi3}. 
The fact that the data can not be described with $s_2 = 0$ leads to 
the interpretation that the elliptic flow is not caused by an 
azimuthal velocity variation alone, but by the combination of a velocity 
difference and a spatially anisotropic freeze-out hyper-surface. 

Table~\ref{fitparameters} lists the results of the two fits (without,
and with $s_2$) for minimum-bias data. 
The parameters are correlated and therefore have large 
uncertainties. 
It is expected that better constraints on the parameters 
can be obtained in the future, when the fits to the single inclusive 
spectra for different mass particles are available. 
As the solid lines in Fig.~\ref{pidflow1} show, the data 
could be described by a reasonable set of parameters. For the 
different centralities, the $T_f$ and $\rho_0$ were kept fixed to 
the minimum-bias values. 
Both the obtained $\rho_a$ and $s_2$, not shown, 
decrease from peripheral to central collisions as expected from a 
simple geometrical picture.

\begin{table}[htb]
  \begin{ruledtabular}
    \begin{tabular*}{\hsize}{m{2.cm}|m{2.cm}|m{2.cm}|m{2.cm}}
      $T_f$(MeV) & $\rho_0$ & $\rho_a$ & $s_2$ \\ 
      \colrule
      135 $\pm$ 19  &  0.58 $\pm$ 0.03 & 0.09 $\pm$ 0.02 & 0 \\
      101 $\pm$ 24  &  0.61 $\pm$ 0.05 & 0.04 $\pm$ 0.01 & 0.04 
      $\pm$ 0.01 \\
    \end{tabular*}
  \end{ruledtabular}
  \caption{The parameters from the Blast Wave fit for minimum-bias 
    collisions.
    The first row lists
    the parameters for the unmodified fit ($s_2 = 0$), the second row
    gives the results from the fit function including the spatially
    anisotropic freeze-out hyper-surface.
    \label{fitparameters}}
\end{table}

A full hydrodynamic calculation, shown in Ref.~\cite{pasi2}, 
describes the $v_2$($p_t$) for pions, kaons and protons + antiprotons for
minimum-bias collisions equally well. 
The agreement is consistent with early local thermal equilibrium and the
presence of an early stage pressure gradient.
However, within the current statistical uncertainties in the data and
different theoretical interpretations~\cite{teaney}, the measurement
does not justify drawing inferences about the different EOS.
Comparing the centrality dependence from our Fig.~\ref{pidflow2} with the
same hydrodynamic calculation, shows a
deviation which is most pronounced for the most peripheral data.
This might indicate that the amount of particle 
rescattering in peripheral collisions is insufficient to justify the 
local thermal approximation implicit in hydrodynamical models.

We have made the first measurement of identified particle elliptic
flow at RHIC.
The measured elliptic flow as a function of $p_t$ and centrality 
differs significantly for particles of different masses.
This mass dependence can be described with a simple 
hydrodynamic-motivated Blast Wave model. 
This model suggests that elliptic flow is generated by the combination of
an azimuthal velocity variation and a spatially anisotropic freeze-out 
hyper-surface.
The mass dependence of $v_2$($p_t$) is also in
close agreement with full hydrodynamic model calculations, suggesting
that in Au+Au collisions at
$\sqrt{s_{_{NN}}}=130$~GeV, a system is created which, for central and
mid-peripheral collisions, is consistent with early
local thermal equilibrium followed by hydrodynamic expansion.

\vspace{0.2cm}

{\bf Acknowledgments:} 
We wish to thank P. Huovinen for providing us with the
hydrodynamical model calculation,
the RHIC Operations Group at Brookhaven National 
Laboratory for their tremendous support and for providing collisions 
for the experiment. This work was supported by the Division of Nuclear 
Physics and the Division of High Energy Physics of the Office of
Science of the U.S.Department of Energy, the United States National 
Science Foundation,
the Bundesministerium f\"{u}r Bildung und Forschung of Germany,
the Institut National de la Physique Nucleaire et de la Physique 
des Particules of France, the United Kingdom Engineering and Physical 
Sciences Research Council, Fundacao de Amparo a Pesquisa do Estado de Sao 
Paulo, Brazil, and the Russian Ministry of Science and Technology.


\begin{thebibliography}{99}
\bibitem{footminbias}
  Minimum-bias corresponds to 0 -- 85\% of the total measured cross
  section, deduced from the measured charged particle 
  multiplicity~\cite{starflow}, excluding the 15\% most peripheral events. 
  The total measured cross section is estimated to correspond to about
  90\% of the geometric cross section, with losses mainly due
  to vertex finding inefficiencies for the low multiplicity events.
\end{thebibliography}
\end{document}